\documentclass[osajnl, 10pt,twocolumn,superscriptaddress,showpacs,amsmath,amssymb,floatfix]{revtex4}
\usepackage{epsfig}
\usepackage{amsmath}
\usepackage{amssymb}
\usepackage{bm}
\usepackage{graphicx}
\usepackage{color}
\usepackage{graphicx,rotating}

\begin{document}
\title{Fields of an ultrashort tightly-focused laser pulse}

\author{Jian-Xing Li}
\affiliation{Max-Planck-Institut f\"{u}r Kernphysik, Saupfercheckweg 1,
69029 Heidelberg, Germany}

\author{Yousef I. Salamin}
\affiliation{Max-Planck-Institut f\"{u}r Kernphysik, Saupfercheckweg 1,
69029 Heidelberg, Germany}
\affiliation{Department of Physics, American University of Sharjah, POB 26666,
Sharjah, United Arab Emirates}

\author{Karen Z. Hatsagortsyan}
\affiliation{Max-Planck-Institut f\"{u}r Kernphysik, Saupfercheckweg 1,
69029 Heidelberg, Germany}

\author{Christoph H. Keitel}
\affiliation{Max-Planck-Institut f\"{u}r Kernphysik, Saupfercheckweg 1,
69029 Heidelberg, Germany}

\date{\today}

\begin{abstract}

Analytic expressions for the electromagnetic fields of an ultrashort, tightly focused, linearly polarized laser pulse 
in vacuum are derived from scalar and vector potentials, using a small parameter which assumes a small bandwidth of the laser pulse.
The derived fields are compared with those of the Lax series expansion and the complex-source-point approaches and  
are shown to be well-behaved and accurate even in the subcycle pulse regime. 
We further demonstrate that terms stemming from the scalar potential and due to a fast varying pulse envelope are non-negligible and may significantly influence laser-matter interactions.

\end{abstract}


\maketitle

\section{Introduction}

With the rapid development of ultrashort and ultrastrong  laser pulses \cite{krausz,wang,eli,piazza_rmp}, laser-matter interactions involving such pulses have been the subject of much interest \cite{brabec2,krushelnick,kitagawa}. Theoretical efforts aimed at describing such pulses have also been evolving to meet the need to model the electric and magnetic fields of the required pulses \cite{horvath,esarey,esarey2,wang1,mora,porras,narozhny,winful,wang2,lu,hua,yan,becker,sepke_pulse,varin,pukhov,April,abdollahpour,marceau,gonoskov,kaertner}

In the study of many laser-matter interactions which employ a  focused laser beam, the laser fields are adequately described by solutions to the wave equation in the paraxial approximation \cite{yariv_book}.
In the case of a tightly focused laser beam, this description is improved upon further using  higher-order corrections to the paraxial solution. For example, in the Lax series solution \cite{lax,davis,barton}, the fields are given in ascending powers of the square of the diffraction angle $\epsilon\equiv w_0/z_r$, and provide a good description for a focused laser beam of long duration. Here $w_0$ and $z_r$ are the waist radius at focus and the Rayleigh length, respectively. It was further shown that high-order corrections to the paraxial solution \cite{hora,sepke} can have non-negligible contributions for evaluations of direct laser acceleration of particles \cite{sal-ol1,sal-apb,salamin-prl}.  

Another solution of Maxwell's equations for a focused laser beam can be obtained using the complex-source-point approach (CSPA) \cite{deschamps,agrawal,couture}. The fields via CSPA, when expanded in terms of the parameter $\epsilon$ near the propagation axis of the focused beam, coincide with those via the Lax series expansion approximation (LSEA) \cite{sal-ol}. For a Gaussian beam it was explicitly shown that the summation of an infinite Lax series reduces exactly to the solution of CSPA \cite{couture}. The  CSPA solution for a propagating spherical wave is singular  at the point $r = \pm i$($z+iz_r$), with the coordinate in the propagation direction $z$ and the radius $r$, which limits its utility far from the propagation axis and causes problems in numerical simulations. However, a method for the regularization of CSPA solutions was found \cite{sheppard,ulanowski}. Rather than a propagating  spherical wave, one may use the field of a simple point source and a point sink placed at the origin, which constitutes a standing spherical wave. Then the focused beam, according to CSPA,  is free from singularities \cite{sheppard,ulanowski,april1,april2,april3,April}. The latter field  was shown  to form  when using a laser beam focusing with a parabolic mirror.  

However, the use of  LSEA, or analogously CSPA, is limited in that the pulse temporal shape is factorized in the field expressions and, consequently, it does not take into account distortions of the pulse during propagation. 
While this may work well for rather long multi-cycle laser pulses, in which the change in pulse-shape is not appreciable, yet for many important current and future developments involving ultrashort and tightly-focused pulses of laser light, the space-time evolution and distortion of the pulse-shape ought to be taken into account \cite{esarey,esarey2,wang1,mora,porras,narozhny,winful,wang2,lu,hua,yan,becker,sepke_pulse,varin,pukhov}. An exact analytical solution of Maxwell's equations is known which describes a focused laser beam of a special form with a cosine-squared pulse shape of arbitrary length \cite{sepke_pulse}. However, this solution in the form of a double sum is inconvenient for analytical treatments.  The 3D vectorial Maxwell's equations are solved in \cite{pukhov} for a laser pulse, yet, only in the paraxial approximation. The extension of CSPA to describe pulsed laser light has also been implemented  \cite{heyman,winful,April,kaertner}. Nonetheless, analytical results have been obtained mostly for isodiffracting laser pulses \cite{winful,marceau}.

The LSEA approach was generalized to describe the spatio-temporal evolution of an ultrashort and tightly-focused laser beam  by solving the wave equation with the use of two small parameters, namely, the spatial diffraction angle $1/k_0 w_0$ and the temporal spreading parameter $1/\omega_0 \tau_0$, where $\tau_0$ is the laser pulse duration, $\omega_0$ the frequency, and $k_0$ the wave number ($\omega_0=ck_0$ in vacuum) \cite{wang2,lu,hua,yan}. The corrections to the paraxial and slowly varying envelope approximations for the fields of the focused radially polarized laser pulse of the $\rm TM_{01}$ Gaussian mode were derived in \cite{varin} with a purely time-domain approach. It was shown that due to those corrections the longitudinal field decreases quickly along the propagation direction.

About two decades ago, Esarey {\it et. al} \cite{esarey} proposed a way to describe the spatio-temporal evolution of an ultrashort and tightly-focused laser beam by solving the wave equation with the use of a single small parameter, namely, the relative bandwidth of the laser pulse  $\varpi\equiv \Delta \omega/\omega_0\ll 1$. 
This solution  describes the field of an ultrashort laser pulse when the duration of the pulse fulfils  the condition $\tau_0\lesssim z_r/c$. The advantage here is that the ultrashort and tightly-focused laser beam is described by a compact expression in terms of an expansion over a single small parameter, in contrast to the approach of \cite{wang2,lu,hua,yan,varin} which uses  expansions over two small parameters. In \cite{esarey}, the laser pulse propagating in a plasma was considered and the transverse fields (with respect to the propagation direction) were calculated from a transverse vector potential alone. However, the description of the longitudinal components of the laser field, which become very significant for tightly focused laser beams \cite{lax,varin}, requires the introduction of a scalar potential. Moreover, when the laser field is described by scalar and vector potentials, the scalar potential will alter the transverse fields, as well.

In this paper we derive analytic expressions for the electromagnetic fields in vacuum of an ultrashort and tightly focused  propagating laser pulse of linear polarization and of a Poisson-like spectrum. The fields are derived using the approach of \cite{esarey}, but combining both  contributions of vector and scalar potentials  to describe properly the longitudinal field of the  ultrashort and tightly focused laser beam. The  field expressions include the first order corrections with respect to the small bandwidth parameter $\varpi$.

The structure of the paper is as follows. In Sec.~\ref{calcul} the calculation of the fields is outlined. Comparison of the obtained field structure with  the CSPA and LSEA solutions for the pulses are discussed in Sec.~\ref{discuss}. And, our conclusions will be given in Sec.~\ref{conclusion}.

\section{Calculation of the fields}\label{calcul}

Expressions for the electromagnetic fields propagating in vacuum may be found from vector and scalar potentials $\bm{A}$ and $\Phi$, respectively, which satisfy the equations

\begin{eqnarray}
\nabla^2\bm{A}-\frac{1}{c^2}\frac{\partial^2\bm{A}}{\partial t^2 }= 0,\\
\nabla^2\Phi-\frac{1}{c^2}\frac{\partial^2\Phi}{\partial t^2 }= 0,
\end{eqnarray}
provided the Lorentz condition (SI units are used),
\begin{eqnarray}
\bm{\nabla}\cdot\bm{A}+\frac{1}{c^2}\frac{\partial\Phi}{\partial t} = 0,
\end{eqnarray}
is satisfied, simultaneously. They are equivalent to the full set of Maxwell's equations in the absence of charges and currents \cite{jackson}. Note that, assuming 
\begin{eqnarray}
\Phi=\phi_0\phi(x,y,z,t) \exp[i(k_0z-\omega_0t)],
\end{eqnarray}
where $\phi_0$ is a constant, and $k_0=2\pi/\lambda_0$ a central wave number, corresponding to the wavelength $\lambda_0$, the Lorentz condition yields
\begin{eqnarray}
\Phi = \frac{c^2\bm{\nabla}\cdot\bm{A}}{i\omega_0 -(\partial\phi/\partial t)/\phi }.
\end{eqnarray}
In the special case of a vector potential polarized linearly, say along the $x$-direction, the vector and scalar potentials satisfy identical (scalar) wave equations. As such, the potentials can only differ by multiplicative constants. We may, therefore, write 
\begin{eqnarray}\label{vector}
\bm{A}=\hat{x}a_0a(x,y,z,t) \exp[i(k_0z-\omega_0t)],
\end{eqnarray}
where $\hat{x}$ is a unit vector in the polarization direction and $a_0$ a constant. Now, setting $\phi=a$ the scalar potential is turned into
\begin{eqnarray}\label{phi}
\Phi = \frac{c^2\bm{\nabla}\cdot\bm{A}}{i\omega_0 -(\partial a/\partial t)/a }.
\end{eqnarray}
Thus, the problem of finding the scalar and vector potentials reduces to finding a solution for 
$a(x,y,z,t)$. Then, expressions for the fields $\bm{E}$ and $\bm{B}$ may, respectively, be found from
\begin{equation}\label{fields}
	\bm{E}=-\frac{\partial\bm{A}}{\partial t}-\bm{\nabla}\Phi, \quad \text{and}\quad \bm{B}=\bm{\nabla}			\times\bm{A}.
\end{equation}
In the absence of a scalar potential function $\Phi$, the electric field $\bm{E}$ will have the same 
polarization as the vector potential $\bm{A}$, whereas the magnetic field $\bm{B}$ will not be altered by introduction of the scalar potential.

Following Esarey {\it et. al} \cite{esarey}, a change of variables is first applied to $\zeta=z-ct$, $\eta=(z+ct)/2$. Consequently, the wave equation, Eq.~(1), becomes 
\begin{eqnarray}\label{new_wave}
\left(\nabla_{\perp}^2+2\frac{\partial ^2}{\partial \eta\partial\zeta}\right)\bm{A}=0.
\end{eqnarray}
Inserting Eq.~(\ref{vector}) into Eq.~(\ref{new_wave}), it reads
\begin{eqnarray}
\left(\nabla_{\perp}^2+2ik_0\frac{\partial}{\partial \eta}+2\frac{\partial^2}{\partial\zeta\partial \eta} \right)a\left(r,\zeta,\eta \right)=0,
\label{eq_a}
\end{eqnarray}    
where $r=\sqrt{x^2+y^2}$. This equation is exact, i.e., no approximations have been assumed yet. When the last term is neglected in Eq.~(\ref{eq_a}), it becomes similar to the paraxial equation \cite{yariv_book}, which has an analytical solution. Using the scaled variables $\tilde{r}=r/w_0$, $\tilde{\eta}=\eta/z_r$, and $\tilde{\zeta}=\zeta/\tau_0$, Eq.~(\ref{eq_a}) reads
\begin{eqnarray}
\left(\tilde{\nabla}_{\perp}^2+4i\frac{\partial}{\partial \tilde{\eta}}+4\varpi \frac{\partial^2}{\partial\tilde{\zeta}\partial \tilde{\eta}} \right)a\left(\tilde{r},\tilde{\zeta},\tilde{\eta} \right)=0.
\label{eq_a_scaled}
\end{eqnarray}
The last term is proportional to a small parameter $\varpi\equiv (\omega_0\tau_0)^{-1}$. Taking into account this term perturbatively, the final solution can be derived as a series with powers of $\varpi$. Most easily the latter can be carried out by  Fourier transforming Eq.~(\ref{eq_a}) with regard to $\zeta$, which yields
\begin{eqnarray}
\label{unphys}
\left[\nabla_{\perp}^2+2i\left(k_0+k\right)\frac{\partial}{\partial \eta}\right]a_k\left(r,k,\eta \right)=0,
\end{eqnarray} 
where
\begin{eqnarray}
\label{ak}
a_k\left(r,k,\eta \right)=\frac{1}{\sqrt{2\pi}}\int_{-\infty}^{\infty}a\left(r,\zeta,\eta \right)\exp(-ik\zeta)d\zeta
\end{eqnarray}
is the Fourier transformation of $a\left(r,\zeta,\eta \right)$. Equation~(\ref{unphys}) has the same mathematical structure as that of the paraxial wave equation, which has explicit solutions for different modes. For instance, taking the lowest-order Gaussian mode, the Fourier transform of the laser envelope reads \cite{lax}
\begin{eqnarray}\label{ak1}
a_k\left(r,k,\eta \right)=f_k\exp\psi_k\left(r,k_0+k,\eta \right),
\end{eqnarray}
with
\begin{eqnarray}
\psi_k\left(r,k_0+k,\eta \right)&=&-\frac{1}{2}\ln(1+\alpha_k^2)-\frac{\rho^2}{1+i\alpha_k}\nonumber \\
                                & &-i\tan^{-1}\alpha_k,
                                \label{psik}
\end{eqnarray}  
where $\alpha_k=\eta/z_{rk}$, $z_{rk}=(k_0+k)w_0^2/2$, $\rho=r/w_0$, $w_0$ is the laser focal radius, and $f_k$ the initial axial envelope profile with the focus taken at $\eta=0$.

The laser pulse envelope $a\left(r,\zeta,\eta \right)$ is calculated as the inverse transform of Eq.~(\ref{ak1}):
\begin{eqnarray}\label{a}
a(r,\zeta,\eta)=\frac{1}{\sqrt{2\pi}}\int_{-\infty}^{\infty}f_k\exp(\psi_k)\exp(ik\zeta)d k.
\end{eqnarray}

The inverse Fourier transform can be carried out analytically if  
$\psi_k(r, k_0+k, \eta)$, viewed as a function of $k'=k+k_0$, is  Taylor-series expanded around the central wavenumber $k_0$, as in
\begin{eqnarray}\label{expansion}
	\psi_k(r,k_0+k,\eta) 
	&=&\sum_{n=0}^\infty\psi_0^{(n)}(r,k_0,\eta)\frac{k^n}{n!},\label{psi}\\
		   	\psi_0^{(n)}(r,k_0,\eta)&\equiv &\frac{\partial^n\psi_0(r,k_0,\eta)}{\partial k_0^n},
\end{eqnarray}
where $\psi_0(r,k_0,\eta)\equiv   \psi_k(r,k_0+k,\eta )|_{k=0}$.  
For instance, the first two terms in the series expansion above are
\begin{eqnarray}
\psi_0^{(0)}&=&-\frac{1}{2}\ln(1+\alpha^2)-\frac{\rho^2}{1+i\alpha}-i\tan^{-1}\alpha,\\
\psi_0^{(1)}&=&\frac{i\alpha}{k_0}\left[\frac{1}{1+i\alpha}-\frac{\rho^2}{(1+i\alpha)^2}\right],\label{psi1}
\end{eqnarray}
where $\alpha=\eta/z_r$, and $z_r$ is the Rayleigh length of the laser beam.

The expansion of Eq.~(\ref{psi}) is valid as long as $k/k_0 \ll 1$. From the uncertainty relation 
$\left(\Delta k\right) \left(\Delta \zeta\right) \sim 1$, with $\Delta k \sim  k$ and $\Delta \zeta \sim c\tau_0  $, one obtains $k\sim 1/c\tau_0$. 
Therefore, the fields derived from this approach are valid when $\tau_0\gg 1/\omega_0$, which is fulfilled for a pulse length of at least half a cycle. Thus, Eq.~(\ref{psi})  represents the sought series expansion over the small parameter $\varpi$.

The expansion Eq.~(\ref{psi}) is applicable in the space-time region where $\alpha \lesssim 1$. Taking into account that $\alpha=\eta/z_r=(z+ct)/2=z/z_r-\zeta/(2z_r)$, the range of applicability of the obtained solution may be estimated by
$z\lesssim z_r$ and $c\tau_0\lesssim z_r$, while the latter reads, equivalently, 
\begin{eqnarray}
\omega_0\tau_0\lesssim (k_0w_0)^2. \label{condition}
\end{eqnarray}

The laser pulse envelope $a(r,\zeta,\eta)$ can, in principle, be obtained by  an inverse Fourier transform, i.e., by inserting Eq.~(\ref{psi}) into Eq.~(\ref{a}).  Furthermore, the fields $\bm{E}$ and $\bm{B}$ may then be calculated via   Eqs.~(\ref{phi}) and (\ref{fields}).

\subsection{Laser pulse with Gaussian spectrum}

For calculating the fields of an ultrashort tightly focused laser pulse one needs to specify the spectral distribution of the field $f_k$ in Eq.~(\ref{a}). Simple analytic expressions will be obtained when the spectrum of the laser pulse  is a Gaussian. Although, in this case it is not possible to avoid unphysical negative and zero frequency components, it still can be applied when the pulse's bandwidth is narrow enough, i.e., $\Delta \omega\ll \omega_0$. More complex solutions based on Poissonian spectra are free from the problem above and discussed in the next section.

For the lowest order solution,  
\begin{eqnarray}\label{a0}
a^{(0)}(r,\zeta,\eta)&=&\exp(\psi_0^{(0)})\frac{1}{\sqrt{2\pi}}\int_{-\infty}^{\infty}f_k\exp(ik\zeta)d k\nonumber\\
&=&\exp(\psi_0^{(0)})f(\zeta),
\end{eqnarray}
since $\psi_0^{(0)}$ is independent of $k$. The laser envelope $f(\zeta)$ is the inverse Fourier transform of $f_k$. The solution of Eq.~(\ref{psik}) is singular at $k=-k_0$. To avoid the physically uninteresting situation arising from $k=-k_0$, when the laser field is axially uniform, see Eq.~(\ref{unphys}), the following Gaussian solution of $f_k$ is considered \cite{esarey}
\begin{eqnarray}\label{fk}
f_k=\frac{\sigma}{k_0}\left(1+\frac{k}{k_0}\right)\exp\left(-\frac{\sigma^2}{2}\frac{k^2}{k_0^2}\right),
\end{eqnarray}
where,  $\sigma=\frac{k_0L}{2\sqrt{2\ln 2}}$, and $L=c\tau_0$ equals the full-width-at-half-maximum (FWHM)  of the laser pulse. The inverse Fourier transform of $f_k$ gives
\begin{eqnarray}\label{gauss1}
f(\zeta)=\left(1+\frac{ik_0\zeta}{\sigma^2}\right)\exp\left(-\frac{k_0^2\zeta^2}{2\sigma^2}\right).
\end{eqnarray}

With Eqs.~(\ref{vector}), (\ref{a0}) and (\ref{gauss1}), the vector potential of the zeroth-order Gaussian solution reads 
\begin{eqnarray}\label{vector2} 
\bm{A}=\hat{x}a_0\beta\left(1+\frac{ik_0\zeta}{\sigma^2}\right)\exp\left(ik_0\zeta-\frac{k_0^2\zeta^2}{2\sigma^2}-\beta\rho^2\right),
\end{eqnarray} 
where we have used the ansatz
\begin{eqnarray}
\exp(\psi_0^{(0)})&=&\beta\exp\left(-\beta\rho^2\right),\\
\beta&=&\frac{1}{1+i\alpha}.
\end{eqnarray}
Following are the electric and magnetic field components, calculated via Eq.~(\ref{fields}) from Eq.~(\ref{vector2}) 
\begin{eqnarray}\label{gaussian0}
E_x&=&\frac{E_1}{C_1} \left[-C_1^2+ \frac{\beta^2x^2}{z_r^2}-\frac{\beta }{k_0 z_r}+\frac{i \beta x^2}{C_1z_r \left(C_2 -\zeta/2\right)^2}\right], \nonumber\\
E_y&=&\frac{E_1 f x y}{C_1 z_r^2}\left[\beta+\frac{i z_r}{C_1(z+iz_r-\zeta/2)^2}\right],\nonumber\\
E_z&=&\frac{E_1 \beta x}{C_1z_r^2}\left[z_r \left(i+\frac{i\beta^2 r^2}{4 z_r^2}+\frac{k_0\zeta}{\sigma^2}\right)-\frac{i \beta}{k_0}\right.\nonumber\\
  && \left. +\frac{i z_r}{\left(\sigma^2-ik_0\zeta\right)}+\frac{C_2z_r}{C_1}\right],\nonumber\\
 B_x&=&0,\nonumber\\
B_y&=&E_1\left(-i+\frac{i \beta}{2 k_0 zr}-\frac{i \beta^2 r^2}{4 z_r^2}-\frac{k_0\zeta}{\sigma^2}+\frac{1}{i\sigma^2+k_0\zeta}\right),\nonumber\\
B_z&=&\frac{E_1 \beta y}{z_r}, 
\end{eqnarray}
where
\begin{eqnarray}
E_1&=&\frac{E_0\beta}{F_N}\left(1+\frac{ik_0\zeta}{\sigma^2}\right)\exp\left(ik_0\zeta+i\varphi_0 -\frac{k_0^2\zeta^2}{2\sigma^2}-\beta\rho^2\right),\nonumber\\
C_1&=&i+\frac{i k_0^2 r^2-C_3}{C_3^2}-\frac{1}{k_0\zeta+i\sigma^2}+\frac{k_0\zeta}{\sigma^2}, \nonumber\\
C_2&=&\frac{C_3-2 i k_0^2   r^2}{C_3^3}+\frac{1}{(k_0\zeta+i\sigma^2)^2}+\frac{1}{\sigma^2}\nonumber\\
C_3&=&2k_0\left(z+iz_r-\zeta/2 \right),
\end{eqnarray}
$\varphi_0$ is the carrier envelope phase and $E_0 = k_0 a_0$ the field amplitude. The various components are scaled to make
$E/E_0=1$, at $x=y=z=t=0$, employing a normalization factor $F_N = i/\left(k_0z_rc_0\right)-ic_0$, with $c_0 = 1+1/\left(2k_0z_r\right)+1/\sigma^2$.

The fields calculated from a vector potential alone \cite{esarey} are clearly transverse, with
\begin{equation}\label{Exx}
	E_x = -E_1 C_1, 
\end{equation}
and $E_y = E_z = 0$. The magnetic field components, however, would be the same as in Eqs.~(\ref{gaussian0}). Thus we see that the scalar potential, which is necessary for describing the longitudinal fields, also modifies the transverse fields and, in this case, the normalization factor is $F_N=i[1+1/(2kz_r)+1/\sigma^2]$.

\subsection{Zeroth-order Poisson solution}

For ultrashort laser pulses the Gaussian spectrum $f_k$  contains unphysical negative frequencies. A spectrum which does not contain negative frequencies has to be chosen to adequately describe ultrashort pulses. Here, we choose the Poissonian \cite{caron,feng}
\begin{eqnarray}\label{poissonk}
f_k=\left(\frac{s}{k_0}\right)^{s+1}\frac{k^s e^{-s k/k_0}}{\Gamma(s+1)}H(k),
\end{eqnarray}
where $s$ is a real positive parameter, $\Gamma (x)$  the gamma function, and $H$($k$) the unit step function which ensures that the pulse does not exhibit negative frequencies. Note that the propagation term $e^{ik\zeta}$ is included therein. The parameter $s$ determines the laser pulse duration $\tau_0$ (the root mean square of the temporal distribution $|f(t)|^2$) via \cite{April}
\begin{eqnarray}
\tau_0=\frac{2s}{\omega_0 \sqrt{2s-1}}.
\end{eqnarray}
Taking  the inverse Fourier transform of $f_k$ given by Eq.~(\ref{poissonk}), the laser envelope is obtained as
\begin{eqnarray}
f(\zeta)=\left(1-\frac{ik_0\zeta}{s}\right)^{-s-1}.
\end{eqnarray} 
With this, the zeroth-order vector potential takes the form 
\begin{eqnarray}\label{vp0}
\bm{A}^{(0)}=\hat{x}a_0\beta\left(1-\frac{ik_0\zeta}{s}\right)^{-s-1}\exp(-\beta\rho^2).
\end{eqnarray}
Finally, the following field components may be obtained
\begin{eqnarray}\label{zeroth-order}
E_x &=& \tilde{E}_1 \left[\frac{2 \beta}{iD_1 w_0^2}\left(\frac{2\beta x^2}{w_0^2}-1-\frac{\beta^2 x^2}{D_1w_0^2 z_r} \right) -iD_1\right],\nonumber\\
E_y &=& \frac{\tilde{E}_1 \beta^2 x y}{iD_1 w_0^4}\left(4-\frac{2\beta}{D_1 z_r}\right),\nonumber\\
E_z &=& \frac{\tilde{E}_1 2\beta x}{D_1 w_0^2}\left(-D_1-\frac{2k_0 s'}{s-ik_0\zeta}-\frac{D_2}{D_1} \right),\nonumber\\
B_y &=& \tilde{E}_1\left(iD_1+\frac{2i k_0s'}{s-ik_0\zeta} \right), \nonumber\\
B_z &=& \frac{2\tilde{E}_1 \beta y}{w_0^2}, 
\end{eqnarray}
where
\begin{eqnarray}
\tilde{E}_1&=&\frac{E_0\beta}{\tilde{F}_N}\left(1-\frac{ik_0\zeta}{s}\right)^{-s-1}\exp(-\beta\rho^2+i\varphi_0),\nonumber\\
D_1 &=& \frac{\beta}{2z_r}-\frac{\rho^2 \beta^2}{2 z_r}-\frac{k_0s'}{s-ik_0\zeta},\\
D_2 &=& -\frac{\beta^2}{4z_r^2}+\frac{\rho^2 \beta^3}{2 z_r^2}+\frac{k_0^2s'}{\left(s-ik_0\zeta\right)^2},
\end{eqnarray}
and, $s'=1+s$. Also, the normalization factor $\tilde{F}_N=-d_0- 2/d_0 w_0^2$, and $d_0=i/(2z_r)-ik_0\left(1+s\right)/s$. 

\subsection{First-order Poisson solution}

The first-order solution of $a(r,\zeta,\eta)$, using Eqs.~(\ref{a}) and (\ref{expansion}), can be written as
\begin{eqnarray}
a(r,\zeta,\eta)&=&\frac{1}{\sqrt{2\pi}}\int_{-\infty}^{\infty}f_k\exp(\psi_0^{(0)}+k\psi_0^{1})\exp(ik\zeta)d k\nonumber\\
&=&\frac{1}{\sqrt{2\pi}}\int_{-\infty}^{\infty}f_k\exp(\psi_0^{(0)})\exp[ik(\zeta-i\psi_0^{(1)})]d k\nonumber\\
&=&f(\zeta)(\zeta-i\psi_0^{(1)})\exp(\psi_0^{(0)}).
\end{eqnarray}
Accordingly, the first-order Poisson-based vector potential $\bm{A}$ reads 
\begin{equation}
A^{(1)} = \hat{x} a_0 \beta\left[1-\frac{ik_0\left(\zeta-i\psi_0^{(1)}\right)}{s}\right]^{-s-1}  \exp(-\beta\rho^2),\label{Apoisson}
\end{equation}
with $\psi_0^{(1)}$ given by Eq.~(\ref{psi1}). Using Eq.~(\ref{Apoisson}), the electromagnetic fields are 
\begin{eqnarray}\label{first-order}
E_x&=&\frac{i\bar{E}_1}{2 z_r}\left(\beta-\rho^2\beta^2+\frac{H_1s'}{H_2s}\right)-\frac{2\tilde{E}_1\beta}{H_3 w_0^2}\left(1+\frac{i\beta s' \alpha}{H_2 s}\right)\nonumber\\
&&+H_4x,\nonumber\\
E_y&=&H_4y,\nonumber\\
E_z&=&\frac{\bar{E}_1 x\beta}{H_3 w_0^2 z_r}\left[2i\beta-i\rho^2\beta^2-\frac{\beta s'(i+3\beta\alpha-\rho^2\beta^2\alpha)}{H_2 s}-\right.\nonumber\\
&&\left. \frac{is'(4kz_r-H_1)}{H_2s}+\frac{\beta(2+s)s'(4kz_r-H_1)\alpha}{H_2^2s^2}\right]-\nonumber\\
&&\frac{\bar{E}_1x\beta^3}{2H_3^2z_r^2w_0^2}\left(1+\frac{i\beta s' \alpha}{H_2s}\right)\left[\frac{s'(2\beta-H_1)(4kz_r-H_1)}{H_2^2s^2\beta^2}\right.\nonumber\\
&&\left. +1-2\rho^2\beta-\frac{2s'(1-2\rho^2\beta-i\beta\alpha+3i\rho^2\beta^2\alpha)}{H_2s}\right],\nonumber\\
B_y&=&\frac{i\bar{E}_1\beta}{2z_r}\left[\rho^2\beta-1+\frac{s'(4kz_r-H_1)}{H_2\beta s}\right],\nonumber\\
B_z&=&\frac{2\bar{E}_1\beta y}{w_0^2}\left(1+\frac{i\beta s'\alpha}{H_2 s}\right),
\end{eqnarray}
where
\begin{eqnarray}
\bar{E}_1 &=&  \frac{E_0 \beta}{\bar{F}_N}\left[1-ik_0\left(\zeta+i\psi_0^{(1)}\right)/s\right]^{-s-1}\times\nonumber\\
    &&  \exp(-\beta\rho^2+i\varphi_0),\\
H_1&=&2kz_r-\beta+\rho^2\beta^2-2i\rho^2\beta^3\alpha+i\beta^2\alpha,\\
H_2&=&1-ik_0\left(\zeta+i\psi_0^{(1)}\right)/s,\\
H_3&=&\frac{\partial a/\partial ct}{a}\nonumber\\
&=&\frac{i\beta}{2z_r}(\rho^2\beta-1)-\frac{iH_1s'}{2H_2z_rs},\\
H_4&=&\frac{4\tilde{E}_1\beta^2 x}{H_3w_0^4}\left[1+\frac{2i\beta s' \alpha^2}{H_2 s}-\frac{\beta^2s'\alpha(2+s)}{H_2^2s^2}-\right.\nonumber\\
&&\left. \frac{\beta(iH_2^2s^2-H_1s'\alpha-iH_2s's-2H_2s's\beta\alpha)}{2H_2^3H_3z_rs^3}\times\right.\nonumber\\
&&\left.(H_2s+is'\beta\alpha)\right],
\end{eqnarray}
$s'=s+1$, and the normalization constant is 
\begin{eqnarray}
\bar{F}_N=\frac{i}{2z_r}\left[1+\frac{s'(2kz_r-1)}{s}\right]+\frac{4iz_rs}{w_0^2(1-2kz_rs')}.
\end{eqnarray}

\begin{figure}[b]
\includegraphics[width=8cm]{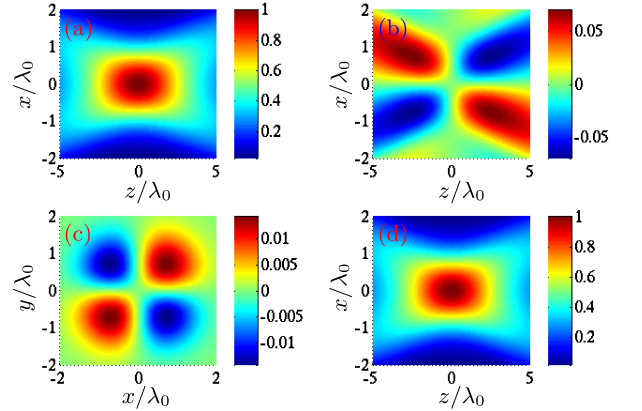}
\caption{(Color online) Contour plots of electric field components normalized by $E_0$ at $t=0$. (a)--(c) are based on the Gaussian field of Eqs.~(\ref{gaussian0}). (a) and (b) show $E_x$ and $E_z$ in the $xz-$plane ($y = 0$) and (c) shows $E_y$ in the $xy-$plane ($z = 0$). (d) Same as (a) but based on the Gaussian field, which is derived solely from the vector potential, see Eq.~(\ref{Exx}). The parameters used are $w_0 = L = \lambda_0 = 1~ \mu$m.}
\label{fig1}
\end{figure}

\section{Discussion}\label{discuss}

Typical distributions of the electric field components are shown in Fig.~\ref{fig1}. Figs.~\ref{fig1}~(a)--(c) are based on Eqs.~(\ref{gaussian0}), derived from combined vector and scalar potentials (VSP) while Fig.~\ref{fig1}(d) is based on Eq.~(\ref{Exx}) or fields derived from a vector potential (VP) alone. Note that Figs.~\ref{fig1}~(a) and (d) are hardly distinguishable for the parameter set used, on account of the fact that the additional terms, except the first one in the $E_x$ component of Eq.~(\ref{gaussian0}), constitute a very small correction to the first term, i.e., the right side of Eq.~(\ref{Exx}). Easily visible deviations may be reached when subcyle pulses are employed instead. On the other hand, $E_z$ and $E_y$, illustrated by Figs.~\ref{fig1}~(b) and (c) are solely due to the VSP approach and are absent from the VP analysis. Those components, although more than two orders of magnitude weaker than their associated $E_x$ component, yet play a significant role in the treatment of interaction of matter with such a laser pulse in vacuum.

\begin{figure}
\includegraphics[width=8cm]{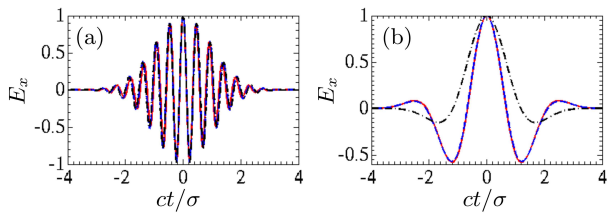}
\caption{(Color online) The time variation of the normalized transverse laser field $E_x$ at focus. The pulse lengths are (a) $L = 5 \lambda_0$ and (b) $L = \lambda_0/2$. The other parameters are the same as in Fig.~\ref{fig1}. The blue dashed curves show our solution  of Eq. (\ref{gaussian0}), while the solid red and dot-dashed black curves are based on fields of  CSPA and LSEA, respectively. Please note that: (a) all three curves are close to each other with merely slight differences; (b) the blue dashed and solid red curves are overlapping. }
\label{fig2}
\end{figure}

Figure \ref{fig2} 
compares and contrasts our Gaussian solutions for  ultrashort focused laser pulses, see Eq.~(\ref{gaussian0}),  with those via CSPA \cite{becker} and LSEA \cite{sal-apb}. In Fig.~\ref{fig2}~(a) the pulse duration is relatively long $L = 5\lambda_0$, while in Fig.~\ref{fig2}~(b) the pulse is of subcyle duration. In both cases the laser beam is tightly focused with a waist size of $w_0=\lambda_0$. It is hard to distinguish the fields stemming from all three approaches for the case of a long laser pulse ($L = 5\lambda_0$). The slight difference between the CSPA and LSEA solutions, on the one hand, and ours, on the other, in this case is due to the fact that the applicability condition Eq. (\ref{condition}) of our solution is at its limit, namely, $\omega\tau_0/(kw_0)^2\approx  0.8$. In the case of a subcycle pulse ($L = \lambda_0/2$), the variations of $E_x$ at focus with time based on the CSPA and our Gaussian solution are almost identical, whereas those based on the LSEA are clearly different due to temporal diffraction effects (the LSEA solution is valid only for pulses of infinitely long duration).

\begin{figure}[h]
\includegraphics[width=7cm]{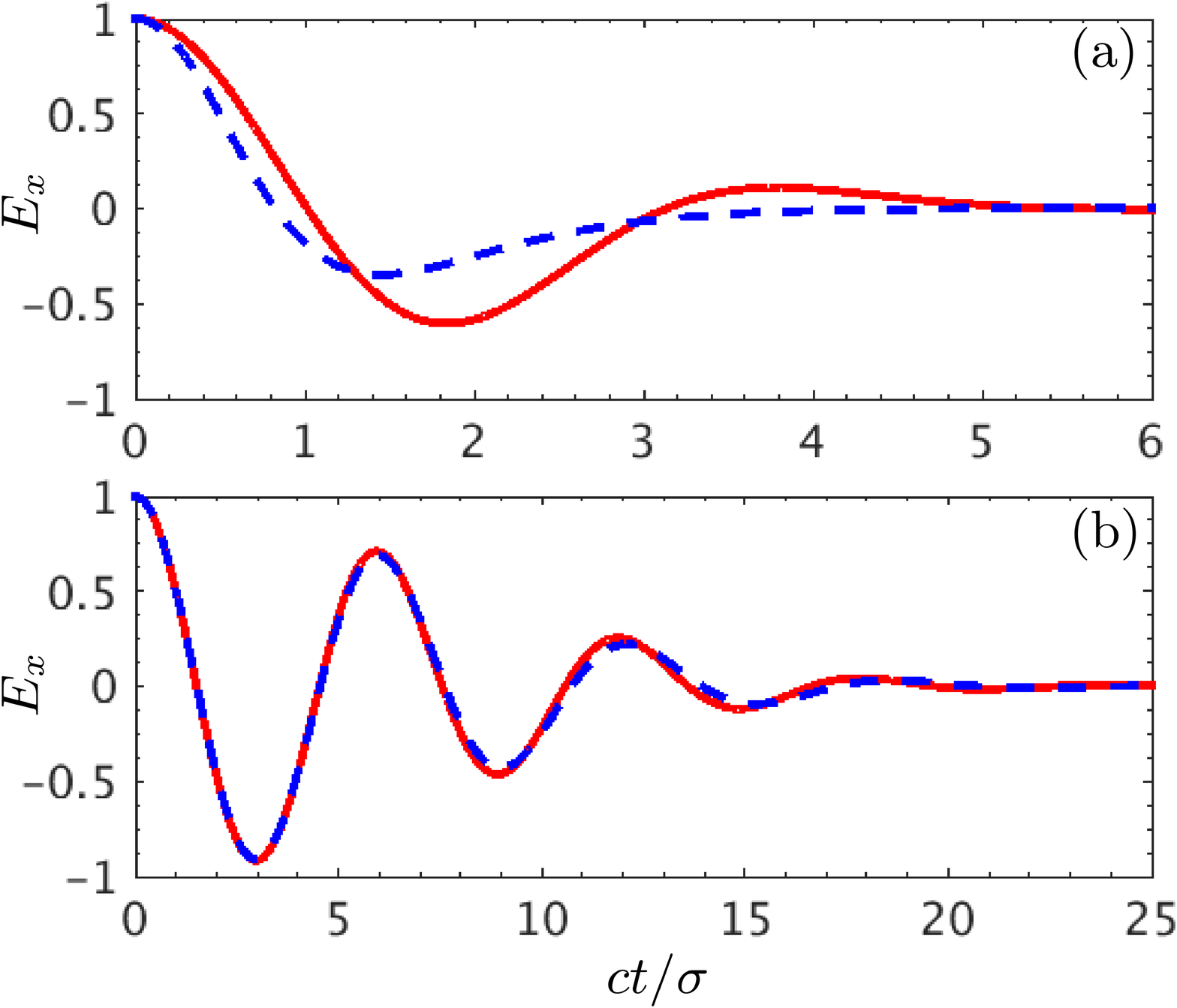}
\caption{(Color online) Time variation of the normalized $E_x$ at focus. The pulse lengths are (a) $L \approx 0.6 \lambda_0$ ($s=2$) and (b) $L \approx 2.7 \lambda_0$ ($s=50$) with $w_0=5\lambda_0$, while the remaining parameters are the same as in Fig.~\ref{fig1}. The red solid and blue dashed curves present the fields of zeroth-order Gaussian and zeroth-order Poisson solution, respectively.  }
\label{fig3}
\end{figure}

In Fig.~\ref{fig3}, we compare the fields of pulses with different spectral distributions, namely, Gaussian and Poissonian. Note that  the red solid and blue dashed curves present the fields of zeroth-order Gaussian and zeroth-order Poisson solution, respectively.  For a subcycle laser pulse of $L\approx 0.6 \lambda_0$ ($s=2$), see Fig.~\ref{fig3}(a), the  field with Gaussian spectrum is significantly different from the Poisson solution. Therefore, the Gaussian spectral profile for the pulse is not suitable anymore, and the Poisson field should be employed to simulate ultrashort pulses.  In Fig.~\ref{fig3}(b), for a several-cycle laser pulse with $L\approx 2.7 \lambda_0$ ($s=50$), the Gaussian's field overlaps with the Poisson solution near the pulse center, and they very slightly differ from each other around the pulse's tail   $ct/\sigma\gtrsim 8$.  Therefore, the Gaussian spectrum is assumed to be an acceptable solution for this case.

\begin{figure}[h]
\includegraphics[width=7cm]{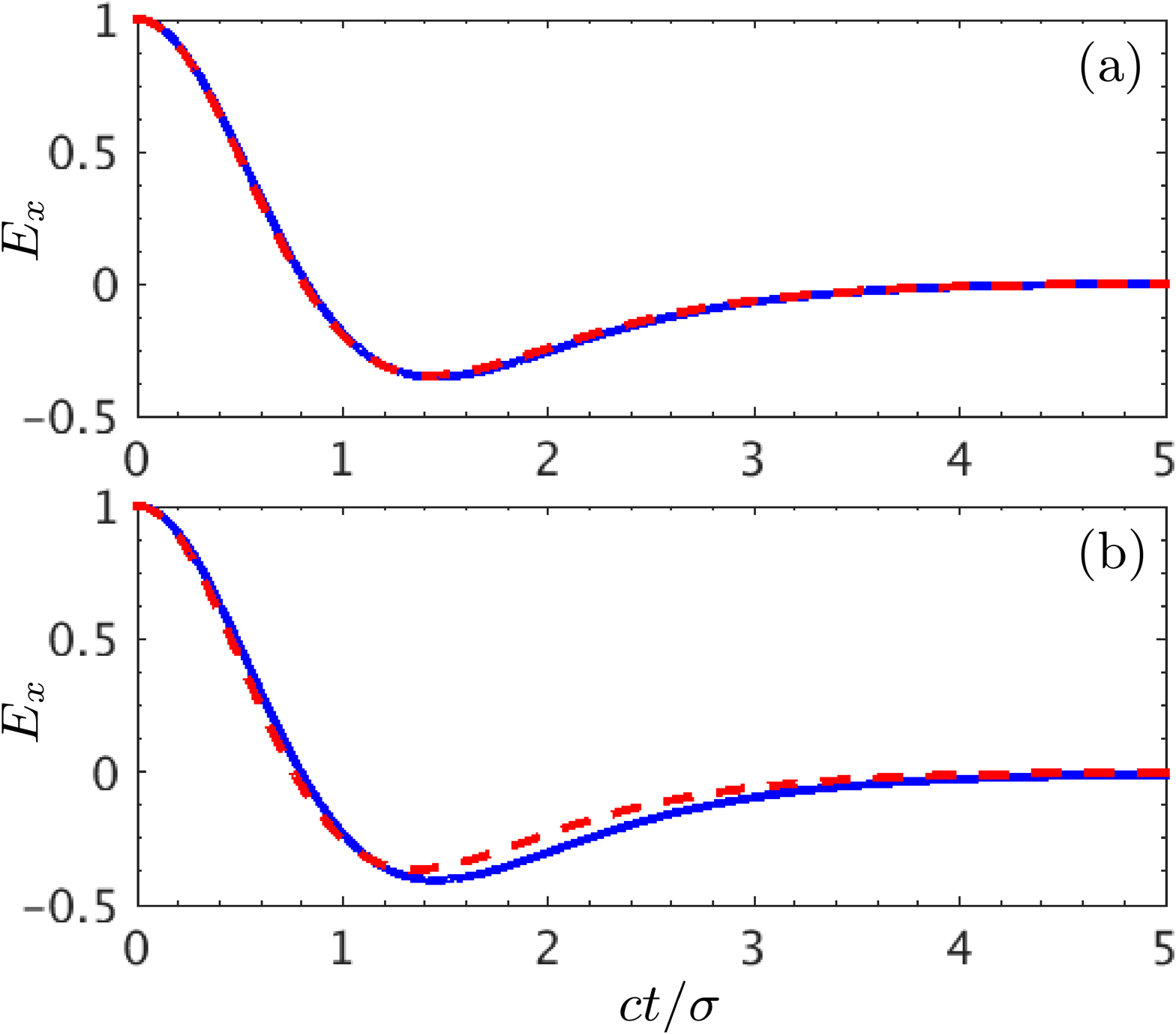}
\caption{(Color online) Time variation of the normalized $E_x$ at focus. (a) $w_0=\lambda_0$, and (b) $w_0=0.5\lambda_0$ with $L \approx 0.6 \lambda_0$ ($s=2$), while the remaining parameters are the same as in Fig.~\ref{fig1}. The blue solid and red dotted curves present the fields of zeroth- and first-order Poisson solutions, respectively.  }
\label{fig4}
\end{figure}

We compare the zeroth- and first-order Poisson solutions for a subcycle laser pulse with different laser focusing sizes in  Fig.~\ref{fig4}. Note that the blue solid and red dotted curves present the fields of zeroth- and first-order Poisson solutions, respectively. For a tightly focused laser pulse  of $w_0=\lambda_0$, see Fig.~\ref{fig4}(a), the first-order solution of the Poisson field overlaps with the zeroth-order solution very well. Therefore, the zeroth-order Poisson field suffices to simulate such pulses. In Fig.~\ref{fig4}(b), for an ultra-tightly focused laser pulse with $w_0=0.5\lambda_0$, the first-order solution of the Poisson field overlaps with the zeroth-order solution near the pulse center as well, however, they are distinguishably different from each other  around the pulse's tail  $ct/\sigma\gtrsim 1$. Consequently, in this case the zeroth-order solution is not sufficient anymore, and the first-order correction should be taken into account.

\section{Conclusion}\label{conclusion}

Following the approach of \cite{esarey}, we have derived the field expressions of an ultrashort tightly focused laser pulse propagating in vacuum, using an expansion over a single small parameter which assumes that the relative bandwidth of the laser pulse is not large. The  basic solution is derived using a Poissonian initial spectral distribution for the pulse, and it is applicable for pulse lengths not larger than the Rayleigh length. The accuracy of the solution is verified in comparison with that of CSPA of \cite{becker}. It is demonstrated, with the introduction of a scalar potential, that the fields do not only develop new axial components, but also modify the transverse ones quite significantly. Moreover, deviations of the derived fields from those based on the LSEA are shown to be non-negligible when the pulse duration is less than a few cycles, and substantial for subcycle pulses.   It has been shown that the Gaussian spectral distribution cannot be applied in the case of  subcycle laser pulses, because they yield pulses differing significantly from those via the Poissonion-spectrum-based ones, with the same bandwidth. Compared with solutions of the wave equation expanded in power-series of two parameters, the diffraction parameter ($1/kw_0$) and the time dilation parameter ($1/\omega\tau_0$), respectively, our solution is more compact, cf., e.g., \cite{yan}, and may be of utility for current and future applications in laser acceleration  and other ultrafast physics. 

\begin{acknowledgments}

YIS has been partially supported by an American University of
Sharjah Faculty Research Grant (FRG-III).

\end{acknowledgments}

\end{document}